\documentclass[openbib,12pt,titlepage,twoside]{article}                                         
\usepackage[widemargins]{a4}                                                                         	   
\usepackage[left=2.5cm,right=2.5cm,top=2.5cm,bottom=3cm]{geometry}
\usepackage{lineno}
\usepackage[onehalfspacing]{setspace}
\usepackage[T1]{fontenc}											
\usepackage{url}   												
\usepackage{graphicx}                                                                                          
\usepackage[margin=35pt,labelfont=bf,textfont=it,font=footnotesize]{caption}     
\usepackage{textcomp}        											
\usepackage{amsmath}                                                                                   	  
\usepackage{amssymb}                                                                                     	  
\usepackage{amsthm}                                                                                       	    
\usepackage{float}                                                                                              	
\usepackage{subfig}   												
\usepackage{booktabs} 
											
\usepackage{natbib}												
\usepackage{here}												
\usepackage{gensymb}													
\usepackage[usenames,dvipsnames,svgnames,table]{xcolor}

\usepackage{lineno}
\begin{document}

\begin{titlepage}
\begin{center}
\setcounter{page}{1}
\Large \textbf{The impact of oceanic heat transport on the atmospheric circulation}\\

\vspace{1.0cm}

\normalsize\textbf{Marc-Andre Knietzsch$^1$, Valerio Lucarini$^{1,2}$ and Frank Lunkeit$^1$}
\vspace{1.0cm}

$^{1}$Meteorologisches Institut, Universit\"at Hamburg, Hamburg, Germany

$^{2}$Department of Mathematics and Statistics, University of Reading, Reading, UK

\vspace{1.0cm}

{\it Correspondence to:} Frank Lunkeit (frank.lunkeit@uni-hamburg.de)

\end{center}

\end{titlepage}

\newpage

\section*{Abstract}
A general circulation model of intermediate complexity with an idealized earthlike aquaplanet setup is used to study the impact of changes in the oceanic heat transport on the global atmospheric circulation. Focus is put on the Lorenz energy cycle and the atmospheric mean meridional circulation. The latter is analysed by means of the Kuo-Eliassen equation.

The atmospheric heat transport compensates the imposed oceanic heat transport changes  to a large extent in conjunction with significant modification of the general circulation. Up to a maximum about 3PW, an increase of the oceanic heat transport leads to an increase of the global mean near surface temperature and a decrease of its equator-to-pole  gradient. For larger transports, the gradient is reduced further but the global mean remains approximately constant. This is linked to a cooling  and a reversal of  the temperature gradient in the tropics. 

A larger oceanic heat transport leads to a reduction of all reservoirs and conversions of the Lorenz energy cycle but of different relative magnitude for the individual components. The available potential energy of the zonal mean flow and its conversion to eddy available potential energy are affected most. 

Both Hadley and Ferrel cell show a decline for increasing oceanic heat transport with the Hadley cell being more sensitive. Both cells exhibit a poleward shift of their maxima, and the Hadley cell  broadens  for larger oceanic transports. The  partitioning by means of the Kuo-Eliassen equation reveals that zonal mean diabatic heating and friction are the most important sources for changes of the Hadley cell while the behaviour of the Ferrell cell is mostly controlled by friction.

\newpage
\section{Introduction}
Astronomical factors and differences of local albedo cause a difference of net incoming shortwave radiation between low and high latitudes leading to differential heating and a surplus of energy in the tropics. 
Considering global and long-term averages, at steady-state the same amount of supplied energy is emitted to space again, and the incoming shortwave radiation is balanced by the outgoing longwave radiation (Peixoto and Oort 1992, Lucarini and Ragone, 2011). 
Atmospheric and oceanic transports, fuelled by instabilities due to the presence of temperature differences related to inhomogeneous heating, tend to reduce such temperature differences, thus acting as a powerful negative feedback that stabilizes the climate system.     

Stone (1978) argued that the magnitude of the total meridional heat transport, i.e. the sum of the oceanic and the atmospheric contributions, is insensitive to the structure and the specific dynamical properties of the atmosphere-ocean system. That is, changes of the oceanic heat transport (OHT) will be compensated by the atmospheric flow and vice versa. In particular, he suggested that the peak of the heat transport is constrained within a narrow range of latitudes regardless of the radiative forcing. If the climate system is at steady-state, the features of the meridional heat transport can be related to the solar constant, the radius of the Earth, the tilt of the Earth's axis and the hemispheric mean albedo. Stone argued that the insensitivity to the structure and the dynamics of the system is due to the correlation of thermal emissions to space, the albedo and the efficiency of the transport mechanisms of the atmosphere and the ocean.      

Enderton and Marshall (2008) discussed the limits of Stone's hypothesis by employing a series of coupled atmosphere-ocean-sea ice model experiments in which the oceanic circulation on an aqua-planet is constrained by different meridional barriers. The presence or absence of the barriers results in significant different climates, in particular in climates with and without polar ice caps. They concluded that Stone's result is a good guide for ice free climates. But, if polar ice caps are present the effect of the related meridional gradients in albedo on the absorption of solar radiation need to be taken into account.       

The atmospheric compensation implies a significant impact of changes in OHT on the atmospheric circulation as a whole which affects the zonally symmetric flow, the zonally asymmetric (eddy) flow and the interplay between both. Thus, changes in OHT have been commonly used to account for paleo-climatic changes (e.g. Rind and Chandler 1991, Sloan et al. 2001, Romanova et al. 2006). Moreover, OHT is an important factor for a potential anthropogenic climate change since significant modifications of it can expected. Unfortunately, large uncertainties exist of changes in the oceanic circulation simulated in climate change scenarios (IPCC 2013). These result from, amongst others, the uncertainties in fresh water forcing due to potential melting of inland ice sheets. To assess the role of the ocean for historical and potential future climates the impact of the OHT on the atmospheric circulation and the underlying mechanisms need to be investigated systematically.

A way of studying the impact of changes in OHT on the atmospheric circulation is to utilize an atmospheric general circulation model coupled to a mixed-layer ocean. In such a model the OHT can be prescribed. Using a present-day setup including continents Winton (2000), Herweijer et al.~(2005) and Barreiro et al.~(2011) found that increasing OHT results in a warmer climate with less sea ice. A reduction of low level clouds and an increase of greenhouse trapping due to a moistening of the atmosphere appeared to be relevant mechanisms. In addition, a weakening of the Hadley cell with increased OHT was found by Herweijer et al.~and Barreiro et al. 

Utilizing an idealized aqua-planet setup Rose and Ferreira (2013) systematically assessed the impact of the OHT on the atmospheric global mean near surface temperature and its equator-to-pole gradient. For warm and ice-free climates they confirm a near-perfect atmospheric compensation of the imposed changes in OHT. Like in the above studies including continents, they found they found an increase in global mean temperature for increasing OHT, accompanied by a decrease in the equator-to-pole temperature gradient. Tropical SSTs showed to be less affected than higher altitudes. The detailed meridional structure of the oceanic heat transport turned out to be less important. Changes in deep moist convection in the mid-latitudes together with an enhanced water vapor greenhouse appear to be the major drivers. Koll and Abbot (2013) confirmed the low sensitivity of tropical SSTs to OHT changes. In their aqua-planet experiments larger OHT leads to a weakening of the Hadley cell which reduces cloud cover and surface winds, and, thus, counteracts surface cooling by increased OHT.

In the present study we extend and supplement the above studies. Based on the experimental setup of Rose and Ferreira our analyses focus on the impact of OHT changes on the atmospheric dynamics. The integrated effect on the atmospheric energetics is assessed by means of the Lorenz energy cycle. Changes in the atmospheric mean meridional circulation are analysed employing the Kuo-Eliassen equation. In Section 2 we describe the model and the experimental design. Section 3 introduces our diagnostics. The results of the analyses are presented in Section 4. A summary and discussion concludes the paper (Section 5).

\section{Model and experimental setup}

The Planet Simulator (PlaSim, \textit{http://www.mi.uni-hamburg.de/Planet-Simul.216.0.html}) is an open source general circulation model (GCM) of intermediate complexity developed at the University of Hamburg. For the atmosphere, the dynamical core is the Portable University Model of the Atmosphere (PUMA) based on the primitive equation multi-level spectral model of Hoskins and Simmons (1975) and James and Gray (1986). 
Radiation is parameterized by differentiating between shortwave and longwave radiation and between a clear or a cloudy atmosphere following the works of Lacis and Hansen (1974) for the short wave part, Sasamori (1968) for the long-wave, and Stephens (1978) and Stephens et al. (1984) for the radiative properties of clouds. Cloud fraction is computed according to Slingo and Slingo (1991). The representation of boundary-layer fluxes and of vertical and horizontal diffusion follows Louis (1979), Louis et al.(1982), Roeckner et al. (1992) and Laursen and Eliasen (1989). The cumulus convection scheme is based on Kuo (1965, 1974). The ocean is represented by a thermodynamic mixed-layer (slab ocean) model including a 0-dimensional thermodynamic sea-ice. 

Following Rose and Ferreira (2013) we use an earthlike aqua-planet setup with zonally symmetric forcing utilizing present day conditions for the solar constant (1365Wm$^{-2}$) and the CO$_2$-concentration (360ppm). The solar insolation comprises an annual cycle but eccentricity is set to zero. Thus, on annual average the forcing is hemispherically symmetric as well. The mixed layer depth is set to $60\,m$.         

A temporally constant flux into the ocean (q-flux) is used to prescribe the oceanic heat transport (OHT) according to the analytic equation given by Rose and Ferreira (2013):

\begin{equation}
\mbox{OHT}=OHT_0 \cdot \sin(\phi)\cos(\phi)^{2N}
\end{equation}

Where $\phi$ denotes the latitude. $N$ is a positive integer which determines the latitude of the maximum of the transport and the shape of its meridional profile, and $OHT_0$ is a constant defining the magnitude. Rose and Ferreira made sensitivity experiments by varying $N$ (ranging from 1 to 8) and by varying the peak transport (ranging from 0PW to 4PW) which is controlled by $OHT_0$. 

\begin{figure}[t]
\centering
\includegraphics[width=0.75\textwidth]{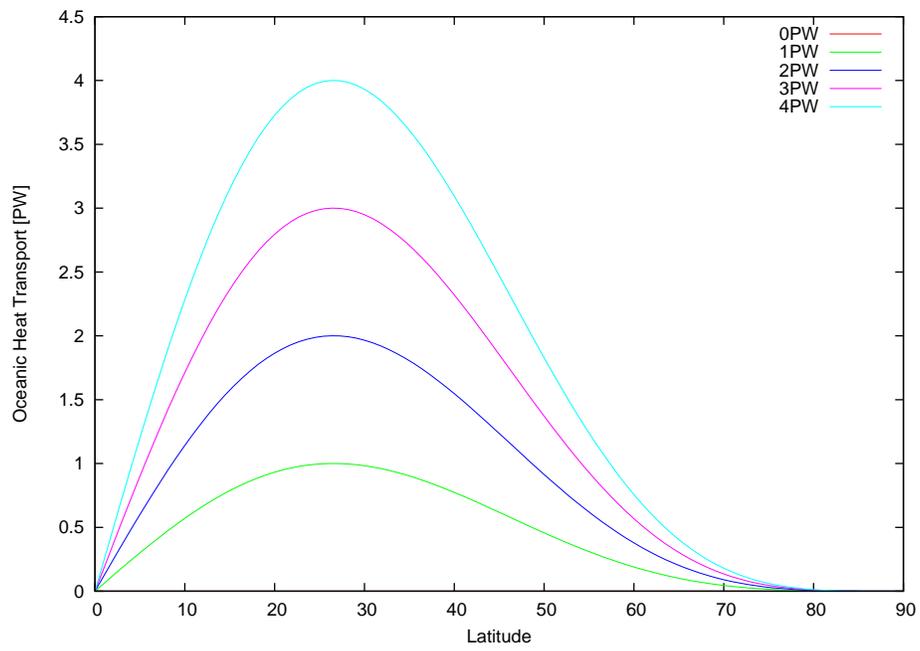}
\centering
\caption{Oceanic heat transport (in PW) for OHT$_{max}$=0,1,2,3, and 4PW.}
\label{fig:OHT}
\end{figure}

For our study we follow Rose and Ferreira but fix the location of the peak by setting $N=2$ (which corresponds to maximum transport at 27$^{\circ}$). We perform nine sensitivity simulations with respect to the magnitude of the transport by changing $OHT_0$ to obtain peak transports OHT$_{max}$ from 0PW to 4PW (with 0.5PW increment). OHT$_{max}$=0PW (i.e. no OHT) serves as the control simulation. The OHT for OHT$_{max}$=0,1,2,3 and 4PW is displayed in Figure \ref{fig:OHT}. 
 
All simulations are run for at least 100 years (360 days per year). The last 30 years are subject to the analyses. A horizontal resolution of $T31$ ($96\,\times\,48$ grid points) with five $\sigma$-levels in the vertical is used. The timestep is $\Delta t=23\,min$.

\section{Diagnostics}

The analyses focus on the Lorenz energy cycle and the mean meridional circulation. The latter will be studied by means of the Kuo-Eliassen equation. Both diagnostics are briefly introduced in the following.

\subsection{The Lorenz energy cycle}
\label{sec:Lorenz}
The atmospheric energy cycle proposed by Lorenz (1955) is one of the most important concepts to understand the global atmospheric circulation by means of energy conservation and by considering the integrated effects of physical mechanisms involved, that is, e.g., the generation of available potential energy by external forcing, the dissipation of kinetic energy and the energy conversions by baroclinic and barotropic processes. At the same time it gives information about the relative importance of the zonal mean circulation, the eddies and the interaction between both. 

Refering to the reservoirs of zonal available potential energy, eddy available potential energy, zonal kinetic energy and eddy kinetic energy as $P_M$, $P_E$, $K_M$ and $K_E$, respectively, the Lorenz energy cycle (i.e. the budget equations) may be written as

\begin{eqnarray*}
\frac{dP_M}{dt} &=& [S_P]-C(P_M,P_E)-C(P_M,K_M) \\
\frac{dP_E}{dt} &=& S_P^*+C(P_M,P_E)-C(P_E,K_E) \\
\frac{dK_E}{dt} &=& S_E^*+C(P_E,K_E)-C(K_E,K_M) \\
\frac{dK_M}{dt} &=& [S_E]+C(K_E,K_M)+C(P_M,K_M) \\
\end{eqnarray*}

where $[S_P]$, $S_P^*$, $[S_E]$ and $S_E^*$ are external sources/sinks of the respective quantities and $C(A,B)$ denotes the conversion from $A$ to $B$. 

To compute the individual contributions we follow the work of Ulbrich and Speth (1991). In pressure coordinates, the reservoirs are given by

\begin{eqnarray*}
P_M &=& \langle \frac{\gamma}{2}([T]-\{T\})^2 \rangle \\
P_E &=& \langle \frac{\gamma}{2}[T^{*2}]\rangle \\ 
K_M &=& \langle \frac{1}{2}([u]^2+[v]^2) \rangle \\ 
K_E &=& \langle \frac{1}{2}([u^{*2}]+[v^{*2}]) \rangle \\
\end{eqnarray*}

and the conversion terms are 

\begin{eqnarray*}
C(P_M,P_E) &=& -\left\langle \gamma \left(
[v^*T^*]\frac{\partial[T]}{r\partial \phi} 
+[\omega^* T^*]\left(
\frac{\partial ([T]-\{T\})}{\partial p}-
\frac{R}{p \cdot cp}([T]-\{T\})\right) \right) \right\rangle \\
C(P_M,K_M) &=& - \left\langle \frac{R}{p}([\omega]-\{\omega\})
([T_v]- \{T_v\}) \right\rangle \\
C(P_E,K_E) &=& - \left\langle \frac{R}{p}[\omega^* T^*_v] \right\rangle \\
C(K_M,K_E) &=& \left\langle \left( [u^*v^*]\frac{\partial[u]}{r \partial \phi} +[u^*v^*][u]\frac{tg \phi}{r}
+[v^*v^*]\frac{\partial[v]}{r \partial \phi} \right. \right. \\
& & \left. \left.
-[u^*u^*][v]\frac{tg \phi}{r}
+[\omega^*u^*]\frac{\partial[u]}{\partial p}
+[\omega^*v^*]\frac{\partial[v]}{\partial p}
\right) \right\rangle \\
\end{eqnarray*}

with 

{\allowdisplaybreaks 
\begin{eqnarray*}
 [x] &=& \mbox{zonal mean} \\
 x^* &=& \mbox{deviation from zonal mean} \\
 \{x\} &=& \mbox{global horizontal mean} \\
 <x> &=& \frac{1}{g\cdot A} \int_A\int_p x dp dA \\
 A &=& \mbox{horizontal Area} \\
 cp &=& \mbox{specific heat at const.~pressure} \\
 g &=& \mbox{gravity} \\
 p &=& \mbox{pressure} \\
 r &=& \mbox{radius of the Earth} \\
R &=& \mbox{gas constant} \\
T &=& \mbox{temperature} \\
T_v &=& \mbox{virtual temperature} \\
u &=& \mbox{zonal wind} \\
v &=& \mbox{meridional wind} \\
\omega &=& \mbox{vertical (p) velocity}\\
\phi &=& \mbox{latitude} \\
\gamma &=& \mbox{stability parameter} = -\frac{R}{p}\left(\frac{\partial [\overline{T}]}{\partial p}
-\frac{R}{cp}\frac{[\overline{T}]}{p}\right)^{-1} \\ 
\end{eqnarray*}
}

The external sources/sinks are diagnosed from the respective residuals. We note that in Ulbrich and Speth this energetics was formulated for mixed space-time domain. In our case, however, the contributions by stationary eddies is zero because of the zonally symmetric forcing. 

\subsection{The mean meridional circulation}

To analyse the mean meridional circulation we make use of the so called Kuo-Eliassen equation (Kuo 1956, Eliassen 1951). It is a diagnostic equation which relates the mean meridional circulation (i.e. Hadley, Ferrel and Polar cell) to different sources. 

Applying the quasi-geostrophic approximation and defining a streamfunction $\psi$ with

\begin{eqnarray*}
 [v] & = & \frac{g}{2\pi r \cos \phi}\frac{\partial \psi}{\partial p} \\
 {[}\omega]  &=& -\frac{g}{2\pi r \cos \phi}\frac{\partial \psi}{r\partial \phi} \\ 
\end{eqnarray*}

the Kuo-Eliassen equation may be derived as (see, e.g., Peixoto and Oort 1992 chapter 14.5.5)

\begin{eqnarray*}
\frac{f^2g}{2\pi r \cos \phi}\frac{\partial^2 \psi}{\partial p^2}
-\frac{g}{2\pi r^3 \rho [\theta ]}\frac{\partial}{\partial \phi}
\left(\frac{\partial [\theta ]}{\partial p}\frac{\partial \psi}{\partial \phi} \right) & = & \frac{1}{r\rho [T]}\frac{\partial}{\partial \phi}\frac{[Q]}{cp} \\
 & - & f \frac{\partial[F]}{\partial p} \\
 & - & \frac{1}{r^2\rho [\theta ]} \frac{\partial}{\partial \phi}
\frac{\partial [v^*\theta^*]\cos\phi}{\cos\phi\partial\phi} \\
 & + & \frac{f}{r\cos^2\phi}\frac{\partial^2[u^*v^*]\cos^2\phi}{\partial p \partial\phi} \\
\end{eqnarray*}  

Where, in addition to the symbols defined above, $f$ is the Coriolis parameter, $\rho$ density, $\theta$ potential temperature, $Q$ diabatic heating and $F$ the tendency of the zonal wind due to friction.

We solve the Kuo-Eliassen equation for $\psi$ by applying an iterative method (Gauss-Seidel method) to its finite difference approximation. Thus, we are able to diagnose the contributions from the different sources to the mean meridional circulation, which are diabatic heating (1$^{st}$ term r.h.s), friction (2$^{nd}$), meridional eddy heat transport (3$^{rd}$) and eddy momentum transport (4$^{th}$). We note that though the equation, in the present form, involves the quasi-geostrophic approximation it has shown to be reasonably applicable even in the deep tropics (Kim and Lee 2001a,b).

\section{Results}
\begin{figure}[t]
\centering
\includegraphics[width=0.75\textwidth]{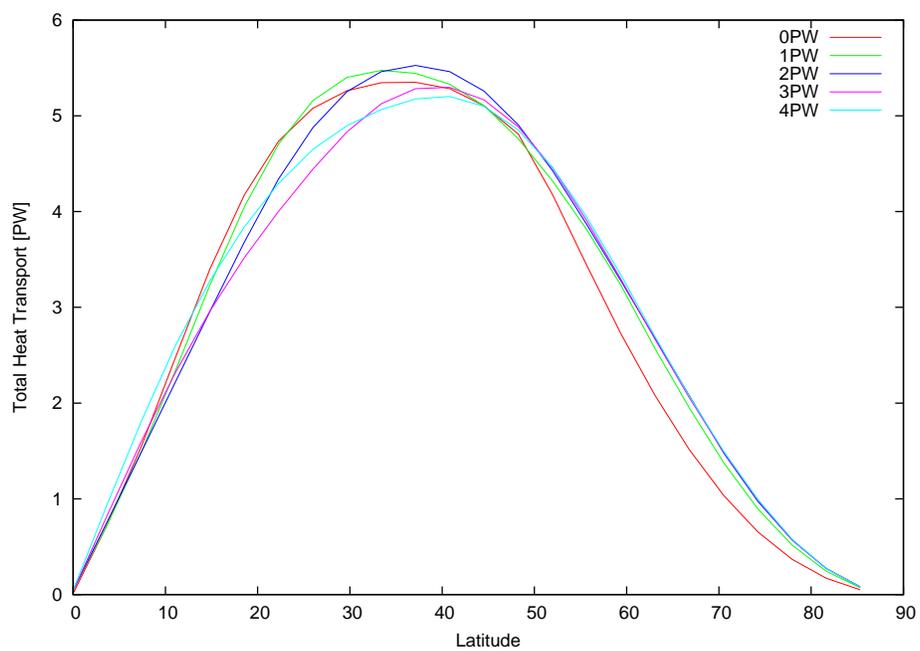}
\centering
\caption{Total heat transport (in PW) diagnosed from energy budget at the top of the atmosphere for OHT$_{max}$=0,1,2,3, and 4PW.}
\label{fig:THT}
\end{figure}

Before discussing the impact of changes in OHT on the Lorenz energy cycle and the mean meridional circulation we present the effect on the mean climate in terms of the total meridional heat transport and the atmospheric near surface (2m) temperature. First we note that similar to Rose and Ferreira our model exhibits multiple equilibria, a warm state and a snow ball Earth depending on the initial conditions as thoroughly discussed in Boschi et al.~(2013). In the present study we investigate the warm states only. However, in contrast to Rose and Ferreira sea ice at high latitudes is present in those warm state in all simulations. Despite the difference in sea ice extend (i.e. planetary albedo) the atmospheric heat transport compensates the changes in OHT to a great extend as can be seen from the total meridional heat transport diagnosed from the energy budged at the top of the atmosphere (Figure \ref{fig:THT}). 

\begin{figure}[t]
\centering
\includegraphics[width=0.75\textwidth]{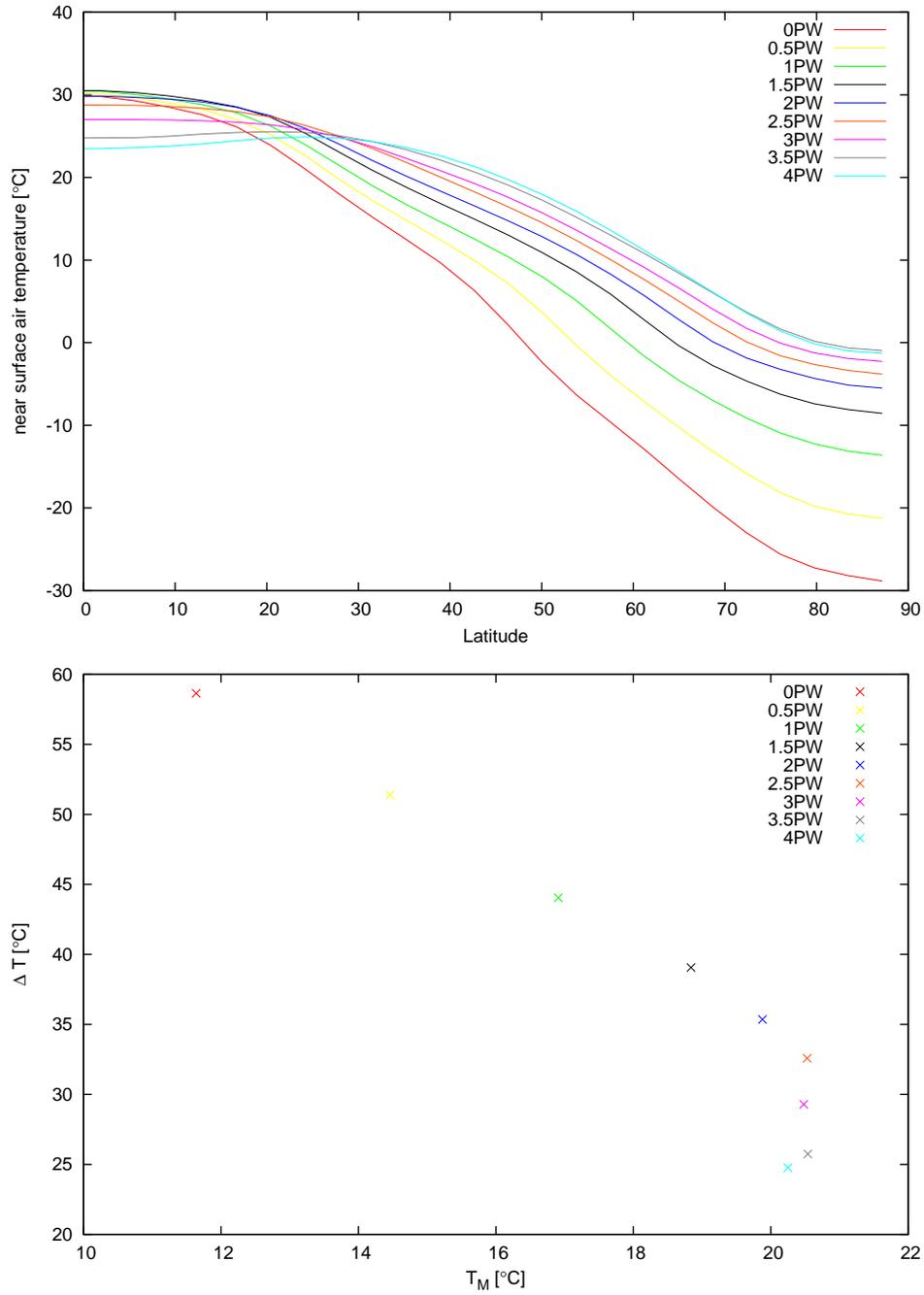}
\centering 
\caption{Climatological annual averages for all simulations: Upper: Zonal mean near surface temperature. Lower: Global mean near surface temperatures (T$_M$,in $^{\circ}$C) versus equator-to-pol gradient ($\Delta T$, in $^{\circ}$C).} 
\label{fig:T2M+TMDT}
\end{figure}

\begin{figure}[t]
\centering
\includegraphics[width=0.75\textwidth]{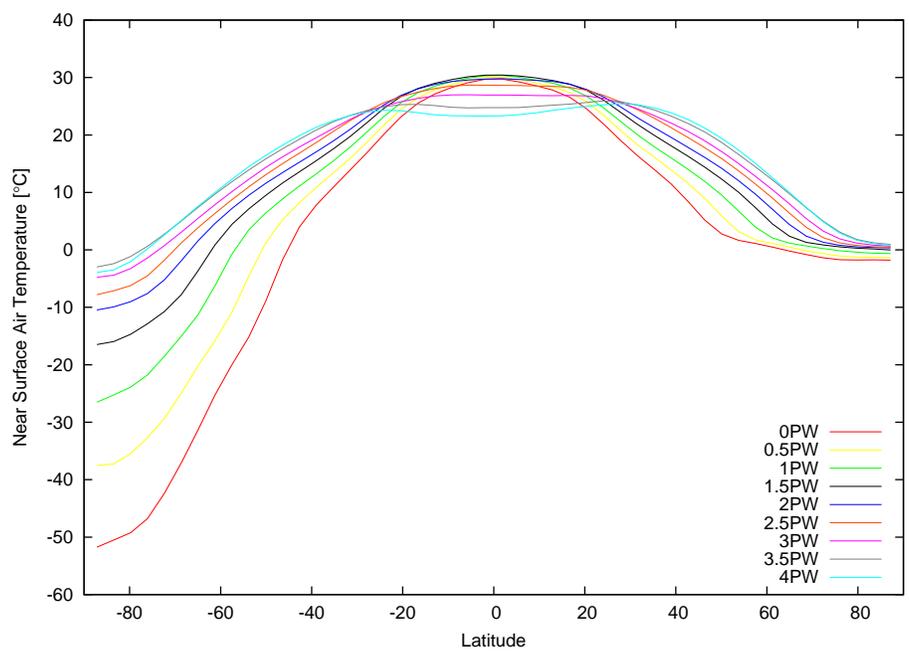}
\centering
\caption{Climatological southern hemisphere summer (June-August) averages for all simulations: Zonal mean near surface temperatures (in $^{\circ}$C).} 
\label{fig:T2MJJA}
\end{figure}

Up to about OHT$_{max}$=2.5PW increasing OHT leads to an increase of the global mean ($T_{M}$) and a decrease of the equator-to-pole gradient ($\Delta T$) of the annual and zonal mean near surface air temperature (Figure \ref{fig:T2M+TMDT}). For this regime an approximately linear relationship between $T_{Mean}$ and $\Delta T$ can be found. For OHT$_{max}$>2.5PW, $T_{M}$ is almost insensitive to an OHT change while $\Delta T$ is further reduced when stronger OHTs are considered. Here, the equator-to-pole gradient is defined by the difference between the values at the lowest and highest latitude of the models grid. Inspecting the respective meridional profiles of the annual and zonal mean near surface temperatures we observe that high latitudes are more sensitive to the OHT changes than low latitudes. With increasing OHT the polar temperatures continuously increase except for OHT$_{max}$=4PW where slightly colder polar temperatures than for OHT$_{max}$=3.5PW are found. It appears that this is a consequence of the reduced atmospheric heat transport slightly over-compensating the increased but still small oceanic heat transport at these latitudes. In the tropics, an increase of the temperatures is only present until OHT$_{max}$=1.5PW. For larger OHT the equatorial temperatures decrease. In addition, increasing OHT leads to a flattening of the temperature profile in the tropics until, for OHT OHT$_{max}$=3.5 and 4PW, the temperature gradient in the tropics gets reversed and the maximum of the temperature shifts away from the equator to approx. $\pm 24\,^{\circ}$. Qualitatively, all findings are also true for winter and summer as can be seen in Figures \ref{fig:T2MJJA}, except that in summer the sensitivity to OHT changes is small in high latitudes which are covered by sea ice. In addition we note that the seasonality and its sensitivity to OHT changes are small for latitudes without sea ice due to the high thermal inertia of the mixed layer. 

\subsection{The Lorenz energy cycle}
\begin{figure}[t]
\centering
\includegraphics[width=0.75\textwidth]{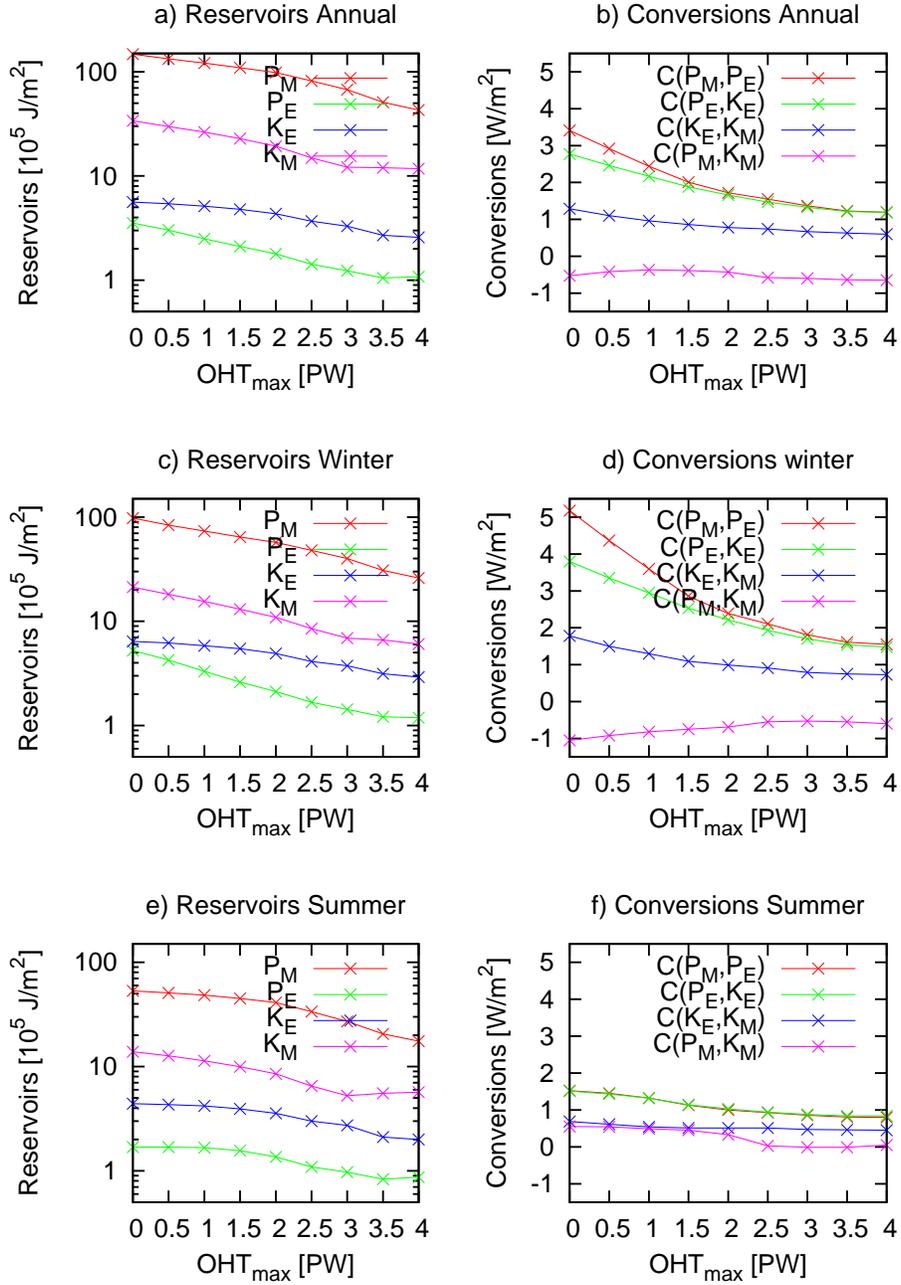}
\centering
\caption{Climatological mean Lorenz energy cycle for June-August: Reservoirs (left, in 10$^5$ J m$^{-2}$) and Conversions (right, in W m$^{-2}$) for yearly data (upper), the winter hemisphere (middle) and the summer hemisphere (lower)for all simulations.} 
\label{fig:LORENZ}
\end{figure}

We compute the climatological average Lorenz energy cycle from 30 year daily data for the entire year and for June-August showing the respective terms for the winter (northern) and the summer (southern) hemisphere separately (Figure \ref{fig:LORENZ}). We note that by using the equations given in Section \ref{sec:Lorenz} the computed averaged values include contributions from the annual cycle. This, however, only affects the annual values of the reservoirs $P_M$ and $K_M$, and of the conversion $C(P_M,K_M)$. Using annual averaged values would diminish the annual $P_M$, $K_M$ and $C(P_M,K_M)$ by about 50\% each. 

The monotonic but nonlinear decrease of the global mean zonal averaged temperature gradient with increasing OHT which is present throughout the troposphere is reflected in the mean flow available potential energy $P_M$ (Figures \ref{fig:LORENZ}a,c,e). Compared to the OHT$_{max}$=0PW simulation, $P_M$ is reduced by about 71\% in the OHT$_{max}$=4PW run for yearly values, 73\% for the winter and 67\% for the summer hemisphere. A decrease with increasing OHT can also be detected for all other reservoirs. However, here the relative decrease for $P_E$ and $K_M$ is substantially smaller for the summer (approx.~48\% and 59\%, resp.) than for the winter hemisphere (approx.~77\% and 72\%). The yearly values reduced by approx.~69\% and 65\%. $K_E$ declines by about 54\% for the yearly data and for both hemispheres, thus showing by far the smallest decrease of all reservoirs on the winter hemisphere. 

The overall decline in the reservoirs with increasing OHT is also present for conversion terms which are mainly related to the activity of baroclinic synoptic eddies, i.e. $C(P_M,P_E)$, $C(P_E,K_E)$ and $C(K_E,K_M)$ (Figures \ref{fig:LORENZ}b,d,f). Here, the largest changes occur in winter where baroclincity (i.e. the temperature gradient) is reduced most. However, the sensitivity appears to decrease following the sequence of a baroclinic live cycle. While $C(P_M,P_E)$ show the largest sensitivity (approx.~65\%, 70\% and 47\% for annual, winter and summer, resp.), the changes in $C(K_E,K_M)$ are the smallest (approx.~53\%, 59\% and 34\%). The changes in $C(P_E,K_E)$ amount to approx.~57\%, 61\% and 44\%,respectively. The convergences of the conversions with increasing OHT indicate that zonally asymmetric diabatic heating and friction become less important for the Lorenz energy cycle. We note that in the present simulations the zonally asymmetric heating extracts $P_E$, i.e.~it acts to homogenize the zonal temperature profiles. For OHT$_{max}$>4PW there is almost no contribution of zonally asymmetric diabatic heating similar to the summer hemisphere in all simulations. $C(P_M,K_M)$ exhibits the smallest sensitivity. For the summer hemisphere, a decrease with increasing OHT can be found indicating that the thermally indirect zonal mean circulation (Ferrel cell) becomes more important compared to the thermally direct one (Hadley Cell). For the winter hemisphere, the thermally direct mean circulation dominates in all simulations but with slightly decreasing magnitude. On annual basis little changes in $C(P_M,K_M)$ are found.      

In summary, an increase in OHT leads to a decrease in all components of the Lorenz energy cycle. This suggests that the atmosphere gets less active and less efficient. We note that the strength of the Lorenz energy cycle can be linked to a Carnot efficiency of the climate system (Lucarini 2009). Indeed, this efficiency declines with increasing OHT. A study on this efficiency, and on other global thermodynamic properties, will be presented in a companion paper. It turns out that changes in the intensity of the water vapor transport are most relevant, similar to Lucarini et al.~(2010).

\subsection{The mean meridional circulation}
\begin{figure}[t]
\centering
\includegraphics[width=0.75\textwidth]{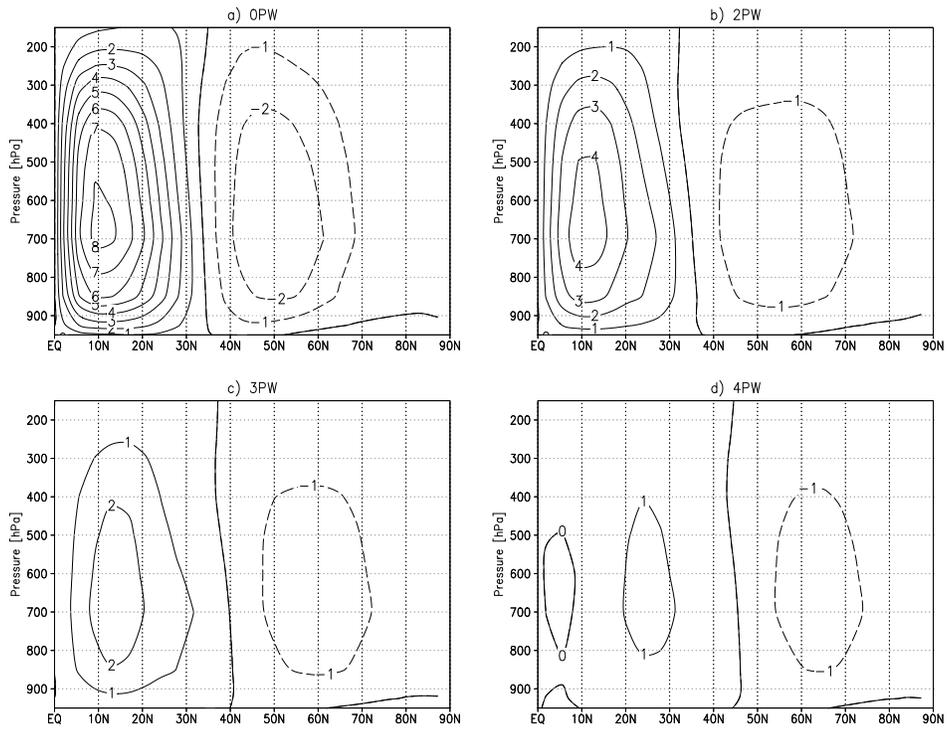}
\centering
\caption{Climatological annual mean mass stream function (in 10$^{10}$ kg s$^{-1}$) for OHT$_{max}$=0,2,3, and 4PW.} 
\label{fig:MMC}
\end{figure}

For discussing the changes in the mean meridional circulation we focus on the climatological annual average of the zonal mean mass stream function ($\psi$). However, if not stated otherwise, the results are qualitatively similar for all seasons. Figure \ref{fig:MMC} shows northern hemisphere $\psi$ for OHT$_{max}$=0, 2, 3 and 4PW. For OHT$_{max}$=0PW, a Hadley cell and a Ferrel cell are well established with values of about 8$\cdot$10$^{10}$kg s$^{-1}$ and -3$\cdot$10$^{10}$kg s$^{-1}$, respectively. The maximum magnitudes are located at about 10$^{\circ}$N for the Hadley cell and 50$^{\circ}$N for the Ferrel cell, and at about 700hPa for both cells. The Hadley cell extends to about 33$^{\circ}$N. A polar cell is absent in the annual mean but emerges weakly in the summer months. 

\begin{figure}[t]
\centering
\includegraphics[width=0.75\textwidth]{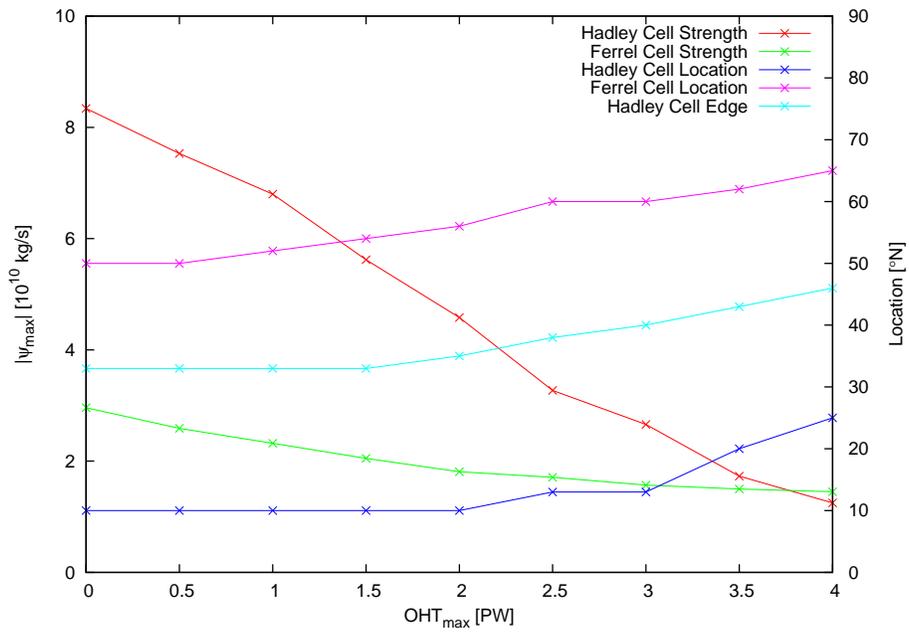}
\centering
\caption{Climatological annual mean mass stream function (northern hemisphere): Strength (in 10$^{10}$ kg s$^{-1}$) and location (in ${^\circ}$N) of Hadley and Ferrel cell for all simulations.} 
\label{fig:HFC}
\end{figure}

With increasing OHT the strength of both cells decreases (Figure \ref{fig:HFC}). However, while the strength of the Hadley cell is virtually linear and amounts to about 85\%, the Ferrel cell strength decreases by about 50\% only with stronger decrease for smaller OHT$_{max}$. In addition to the changes in strength, the Ferrel cell shifts poleward. For OHT$_{max}$>2PW a poleward shift can also be observed for the Hadley cells maximum together with a broadening of this cell, i.e.~a poleward shift of its edge. For OHT$_{max}$=4PW an additional thermally indirect cell can be observed close to the equator. This is related to an almost vanished Hadley cell in summer together with a winter hemisphere Hadley cell which has its maximum on the summer hemisphere. 

\begin{figure}[t]
\centering
\includegraphics[width=0.75\textwidth]{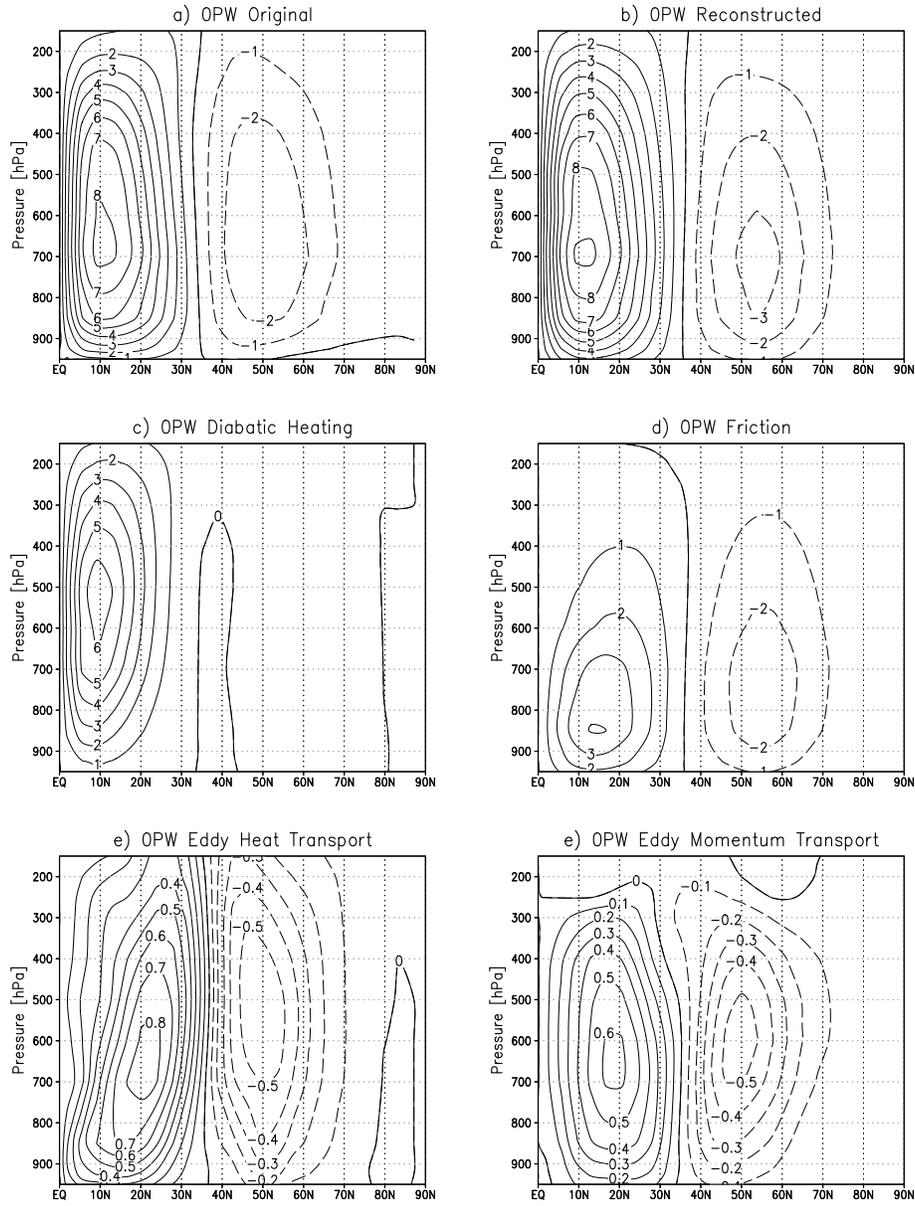}
\centering
\caption{Climatological annual mean mass stream function (in in 10$^{10}$ kg s$^{-1}$) for OHT$_{max}$=0PW: a) Original (see Figure \ref{fig:MMC}); b) Computed from The Kuo-Eliassen equation (all sources); c) Source from diabatic heating; d) Source from friction; e) Source from eddy heat transport; f) Source from eddy momentum transport.} 
\label{fig:MMCR}
\end{figure}

The reconstructions by means of the Kuo-Eliassen equation are in good agreement with the actual $\psi$ for all simulations even if the maximum magnitudes are somewhat overestimated. This gives us confidence to the applied methodology. Figure \ref{fig:MMCR} shows the sources and the reconstruction for OHT$_{max}$=0PW. The individual sources indicate that the largest contributions to $\psi$ stem from diabatic heating and from friction. The heating controls the Hadley cell together with a significant contribution by friction. For the Ferrel cell friction is most important. To a much smaller extent the eddy transports of heat and of momentum add to the Ferrel circulation with a larger contribution by the heat transport. For both Hadley and Ferrel cell the maximum contribution by friction is located at lower levels than for all other sources. 

\begin{figure}[t]
\centering
\includegraphics[width=0.75\textwidth]{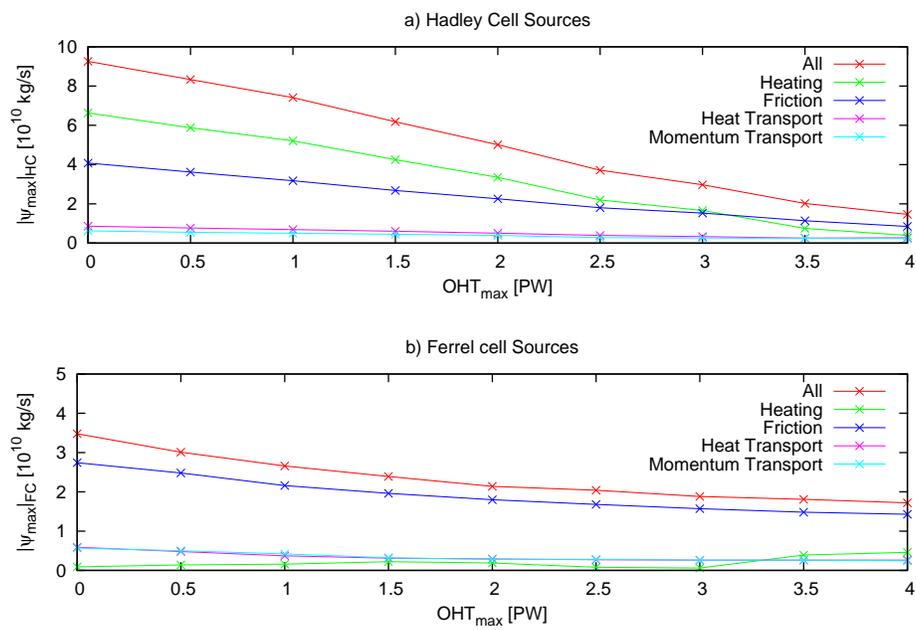}
\centering
\caption{Sources (in 10$^{10}$ kg s$^{-1}$) of the Hadley (upper) and the Ferrel (lower) cell according to the Kuo-Eliassen equation.} 
\label{fig:MMCF}
\end{figure}

For the Hadley cell both major sources, heating and friction, decrease linearly with increasing OHT (Figure \ref{fig:MMCF}). As the decrease is stronger for heating, friction becomes the major source for OHT$_{max}$>3PW. The decrease of the Ferrel cell with increasing OHT is linked to a decrease of the friction, i.h. a decrease of the near surface zonal mean zonal wind. The contributions by the heat and momentum transports decrease less and remain constant for OHT$_{max}$>2PW. Similar to the changes of magnitude, the shifting of the cells and the broadening of the Hadley cell can be explaind by respective changes in the mean sources.

\section{Summary and discussion}
We have studied the impact of the oceanic heat transport (OHT) on the atmospheric circulation focusing on the Lorenz energy cycle and the mean meridional circulation. Utilizing a general circulation model of intermediate complexity (PlaSim) which includes an oceanic mixed layer we have adopted an experimental design from Rose and Ferreira (2013). Here, an imposed oceanic heat transport of simple analytic form and with varying strength allows for systematic analyses. The meridional circulation has been studied by means of the Kuo-Eliassen equations which separates contributions by different sources: diabatic heating, friction and the eddy transport of heat and momentum.

We found a compensation of the changes in oceanic heat transport by the atmosphere confirming Stones (1978) conclusions. The presence of sea ice may explain the deviations from a perfect compensation as discussed in Enderton and Marshall (2008).  
 
In agreement with Rose and Ferreira we have found an increase of the global mean near surface temperature and a decrease of the equator-to-pole temperature gradient with increasing OHT for OHT$_{max}$<3PW. For larger OHT the temperature gradient is still decreasing but the global average remains constant. For the tropics, there is a significant decrease of both temperature and its gradient for OHT$_{max}$>2PW with a reversal of the gradient for OHT$_{max}$>3PW. For smaller OHT we observed a slight warming and a reduction of the gradient with increasing OHT which is consistent with results from Koll and Abbot (2013). However, in their aqua-planet the tropical temperature increases for all imposed (positive) OHTs (up to 3PW). A tropical cooling for imposed oceanic heat transports somewhat larger than present-day values has also been found by Barreiro et al.~(2011) in a more complex coupled atmosphere-slab ocean model with present-day land-sea distribution suggesting that present-day climate is close to a state were the warming effect of OHT is maximized. Barreiro et al.~related the tropical cooling to a strong cloud-SST feedback and showed that the results are sensitive to the particular parameterizations. 

With respect to the Lorenz energy cycle, the decline of the meridional temperature gradient with increasing OHT is directly linked to a reduction of the available potential energy of the zonal mean flow ($P_M$). In addition, the reduced gradient indicates less baroclinicity as well as a weaker zonal mean zonal wind (jet) due to the the thermal wind relation. Finally, less baroclinicity hint to a reduced eddy activity. Thus, it is consistent that all reservoirs and conversions of the Lorenz energy cycle decrease with increasing OHT. However, the sensitivities differ. $P_M$ and the conversion from $P_M$ to $P_E$ exhibit the largest changes. Eddy kinetic energy, the barotropic conversion from eddy kinetic energy to zonal mean kinetic energy, and the conversion from zonal mean kinetic energy to $P_M$ are least affected. 

Confirming the results of previous studies (Herweijer et al.~(2005), Barreiro et al., Koll and Abbot) we have found a decrease of the Hadley cell for increasing OHT. In addition, the Hadley cell broadens and its maximum shifts polward when OHT obtains large values (OHT$_{max}$>2.5PW). Separating individual sources by applying the Kuo-Eliassen equation showed that the characteristics of the Hadley cell can be explained by the mean meridional circulations related to the diabatic heating and, to a smaller extend, to the friction. In our simulations, the meridional circulation induced by friction also controls the behavior of the Ferrel cell. Eddy transports of heat and momentum appear to be less important. This is different from results by Kim and Lee (2001b) where the mean meridional circulation related to eddy fluxes account for about 50\% of the Ferrel cells strength. The coarse vertical resolution adopted from Rose and Ferreira may be responsible for a reduced eddy activity.   

Overall our study demonstrate the large impact of the oceanic heat transport on the atmospheric circulation which effects the zonally symmetric flow, the zonally asymmetric flow and the interaction between both. By reducing the meridional temperature gradient an increased oceanic heat transport weakens the Lorenz energy cycle and slows down and shifts the Hadley and the Ferrel cell. 

The reduction of the meridional gradient of the surface temperature is one of the major features of global warming. Lu et al. (2007) showed a consistent weakening and poleward expansion of the Hadley cell in IPCC AR4 simulations. Hence, changes in the oceanic heat transport may significantly modify the response of the atmospheric circulation to greenhouse warming. A weakening of the oceanic meridional overturning circulation as predicted by the majority of coupled ocean-atmosphere general circulation models (though with large uncertainties; IPCC 2013) would therefor act as a negative feedback mechanism. 
This negative feedback might become even more important when strong melting of inland ice sheets due to global warming is taken into account. The associated input of large amounts of fresh water has a huge potential to slow down the oceanic circulation. 
 
Finally, all changes discussed above can also be seen as a phenomenological fingerprint of modified global thermodynamic properties of the atmospheric flow like entropy production and an Carnot efficiency (Lucarini 2009). Analysing these properties will provide a more general perspective in the context of geophysical fluid dynamics. Results from such an investigation will be presented in a companion work (Schr\"oder et. al. 2014).

\section*{References}

Barreiro, M., Cherchi, A., and Masina, S.: Climate sensitivity to changes in ocean heat transport, J.~Climate, 24, 5015-5030,2011.

Boschi, R., Lucarini, V., and Pascale, S.: Bistability of the climate around the habitable zone: a thermodynamic investigation, Icarus, 226, 1724-1742, 2012.

Eliassen, A.: Slow frictionally controlled meridional circulation in a circular vortex, Astro.~Norv., 5, 19-60, 1951.

Enderton, D.~and Marshall, J.: Explorations of atmosphere-ocean-ice climates on an aquaplanet and their meridional energy transports, J.~Atmos.~Sci., 66, 1593-1610, 2008.

Herweijer, C., Seager, R., Winton, M., and Clement, A.: Why ocean heat transport warms the global mean climate, Tellus, 57A, 662-675, 2005.

Holton, J. R., 1992: An Introduction to Dynamic Meteorology, International Geophysics Series, 48(3), Academic Press, 1992.

Hoskins, B.J.~and Simmons, A.J.: A multi-layer spectral method and the semi-implicit method, Quart.~J.~Roy.~Meteorol.~Soc., 101, 637-655, 1975.

IPCC: Climate Change 2013: The Physical Science Basis. Contribution of Working Group I to the Fifth Assessment Report of the Intergovernmental Panel on Climate Change [Stocker, T.F., D.~Qin, G.-K.~Plattner, M.~Tignor, S.K.~Allen, J.~Boschung, A.~Nauels, Y.~Xia, V.~Bex and P.M.~Midgley (eds.)]. Cambridge University Press, Cambridge, United Kingdom and New York, NY, USA, 1535pp, 2013.

James I.N.~and Gray, L.J.: Concerning the effect of surface drag on the circulation of a planetary atmosphere, Quart.~J.~Roy.~Meteorol.~Soc., 112, 1231-1250, 1986.

Kim, H.-K.~and Lee, S.: Hadley cell dynamics in a primitive equation model. Part I: Axisymmetric flow, J.~Atmos.~Sci., 58, 2845-2858, 2001a.

Kim, H.-K.~and Lee, S.: Hadley cell dynamics in a primitive equation model. Part I: Nonaxisymmetric flow, J.~Atmos.~Sci., 58, 2859-2871, 2001b.

Kuo, H.-L.: Forced and free meridional circulations in the atmosphere, J.~Meteor., 13, 561-568, 1956.

Kuo, H.-L.: On formation and intensification of tropical cyclones through latent heat release by cumulus convection, J.~Atmos.~Sci., 22, 40-63, 1965.

Kuo, H.-L.: Further studies of the parameterization of the influence of cumulus convection on large-scale flow, J.~Atmos.~Sci., 31, 1232-1240, 1974.

Koll, D.D.B.~and Abbot, D.S.: Why tropical sea surface temperature is insensitive to ocean heat transport changes, J.~Climate, 26, 6742-6749, 2013.

Lacis, A.A.~and Hansen, J.E.: A parameterization for the absorption of solar radiation on the Earth's atmosphere, J.~Atmos.~Sci., 31, 118-133, 1974.

Laurson, L.~and Eliasen, E.: On the effects of the damping mechanisms in an atmospheric general circulation model, Tellus, 41A, 385-400, 1989.

Lorenz, E.N.: Available potential energy and the maintenance of the general circulation, Tellus, 7, 157-167, 1955.

Louis, J.F.: A parametric model of vertical eddy fluxes in the atmosphere, Boundary Layer Meteorology, 17, 187-202, 1979.

Louis, J.F., Tiedtke, M., and Geleyn, M.: A short history of the PBL parameterisation at ECMWF, Proceedings, ECMWF workshop on planetary boundary layer parameterization, Reading, 25-27 Nov. 81, 59-80, 1982.

Lu, J., Vecchi, G.A., and Reichler, T.: Expansion of the Hadley cell under global warming, Geophys.~Res.~Lett., 34, L06805, 2007.

Lucarini, V.: Thermodynamic efficiency and entropy production in the climate system, Phys.~Rev.~E, 80, 021118, 2009.

Lucarini, V., Fraedrich, K., and Lunkeit, F.: Thermodynamics of climate change: generalized sensitivities, Atmos.~Chem.~Phys., 10, 9729-9737, 2010.

Lucarini, V.~and Ragone, F.: Energetics of climate models: Net energy balance and meridional enthalpy transport, Rev.~Geophys., 49, RG1001, doi:10.1029/2009RG000323, 2011.

Peixoto, J.P.~and Oort, A.H.: Physics of Climate, American Institute of Physics, 1992.

Rind, D.~and Chandler, M.: Increased ocean heat transports and warmer climate, J.~Geophys.~Res., 96, 7437-7461, 1991.

Roeckner, E., Arpe, K., Bengtsson, L., Brinkop, S., D\"umenil, L., Esch, M., Kirk, E., Lunkeit, F., Ponater, M., Rockel, B., Sausen, R., Schlese, U., Schubert S., and Windelband, M.: Simulation of the present-day climate with the ECHAM-3 model: Impact of model physics and
resolution, Max-Planck Institut f\"ur Meteorologie, Report No. 93, 171pp, 1992.

Romanova, V., Lohmann, G., Grosfeld, K., and Butzin, M: The relative roles of oceanic heat transport and orography on glacial climate, Quat.~Sci.~Rev., 25, 832-845, 2006.

Rose, B.~and Ferreira, D.: Ocean heat transport and water vapor greenhouse in a warm equable climate: a new look at the low gradient paradox, J.~Climate, 26, 2117-2136,doi:http://dx.doi.org/10.1175/JCLI-D-11-00547.1, 2013.

Sasamori, T.: The radiatice cooling calculation for application to general circulation experiments, J.~Appl.~Meteor., 7, 721-729, 1968.

Slingo, A.~and Slingo, J.M.: Response of the National Center for Atmospheric Research community climate model to improvements in the representation of clouds, J.~Geoph.~Res., 96, 341-357, 1991.

Sloan, L.C., Walker, J.C.G., and Jr, T.C.M.: Possible role of oceanic heat transport in early Eocene climate, Paleoceanography, 10, 347-356, 1995.

Schr\"{o}der, A., Lucarini, V., and Lunkeit, F.: The impact of oceanic heat transport on the atmospheric circulation: a thermodynamic perspective, in prep., 2014.

Stephens, G.L.: Radiation profiles in extended water clouds. II: Parameterization schemes, J.~Atmos.~Sci., 34, 2123-2132, 1978.

Stephens, G.L., Ackermann, S., and Smith, E.A.: A shortwave parameterization revised to improve cloud absorption, J.~Atmos.~Sci., 41, 687-690, 1984.

Stone, P. H.: Constraints on dynamical transports of energy on a spherical planet, Dyn.~Atmos.~Oceans, 2, 123-139, 1978.

Ulbrich, U.~and Speth, P.: The global energy cycle of stationary and transient atmospheric waves: Results from ECMWF analyses, Meteorol.~Atmos.~Phys., 45, 125-138, 1991.

Winton, M.: On the climate impact on ocean circulation, J.~Climate, 16, 2875-2889, 2003.

\clearpage

\clearpage

\clearpage

\clearpage

\clearpage
\clearpage
\clearpage
\clearpage
\clearpage
\end{document}